\documentclass[runningheads]{llncs}
\usepackage{array}
\usepackage{booktabs}
\usepackage{rotating}
\usepackage{graphicx}
\usepackage{tabularx}
\usepackage{longtable}
\usepackage{adjustbox}
\usepackage{caption}
\usepackage{makecell}
\makeatletter
\renewcommand\subsubsection{\@startsection{subsubsection}{3}{0mm}%
                                {0.5ex plus 0.1ex minus 0.1ex}%
                                {-1em}
                                {\normalfont\normalsize\bfseries\textbf{}}}
\makeatother

\begin{document}

\titlerunning{Generative AI in Computer Science Education} % Shorter title

\title{A Review of Generative AI in Computer Science Education: Challenges and Opportunities in Accuracy, Authenticity, and Assessment\thanks{This paper was submitted under the Research Track on Education (CSCI-RTED). This material is based upon work supported by the National Science Foundation under Grant No. 2142503.}}

\author{Iman Reihanian\inst{1} \and Yunfei Hou\inst{2} \and Yu Chen\inst{3} \and Yifei Zheng\inst{4}}
\authorrunning{I. Reihanian et al.}
\institute{%
  School of Computer Science and Engineering, California State University, San Bernardino, USA\\
  \email{iman.reihanian1166@coyote.csusb.edu}
  \and
  School of Computer Science and Engineering, California State University, San Bernardino, USA\\
  \email{hou@csusb.edu}
  \and
  School of Information Systems and Technology, San Jose State University, USA\\
  \email{yu.chen@sjsu.edu}
  \and
  University of California, San Diego, USA\\
  \email{yiz152@ucsd.edu}
}

\maketitle

\begin{abstract}

This paper surveys the use of Generative AI tools, such as ChatGPT and Claude, in computer science education, focusing on key aspects of accuracy, authenticity, and assessment. Through a literature review, we highlight both the challenges and opportunities these AI tools present. While Generative AI improves efficiency and supports creative student work, it raises concerns such as AI hallucinations, error propagation, bias, and blurred lines between AI-assisted and student-authored content. Human oversight is crucial for addressing these concerns. Existing literature recommends adopting hybrid assessment models that combine AI with human evaluation, developing bias detection frameworks, and promoting AI literacy for both students and educators. Our findings suggest that the successful integration of AI requires a balanced approach, considering ethical, pedagogical, and technical factors. Future research may explore enhancing AI accuracy, preserving academic integrity, and developing adaptive models that balance creativity with precision.

\keywords{Generative AI \and Computer Science Education \and AI Accuracy \and AI Authenticity \and AI Assessment}
\end{abstract}

\section{Introduction}
With the rise of ChatGPT and large language models, a student in today’s classroom can sit at their computer, not just writing code but co-developing it with an AI partner—a virtual assistant that suggests code snippets, refines algorithms, and troubleshoots errors. The lines between what the student creates and what the AI generates become blurred. How, then, do we ensure that the student is truly mastering the core concepts of programming? Are the algorithms being created as accurate as they appear? And most critically, how do we assess a student’s progress in an educational environment where AI can both aid and overshadow the human creative process? These questions represent real challenges in today’s computer science education, where concerns about AI’s accuracy, authenticity, and assessment are no longer just emerging but are core issues shaping the integration of generative AI (GenAI) into the classroom.

Unlike traditional AI systems that analyze existing data to predict outcomes, Generative AI uses deep learning algorithms to generate new content, such as text, images, and code, to assist and enhance educational activities. By leveraging the ability of GenAI to generate, refine, and troubleshoot code, students can use AI models to explore creative solutions, receive instant feedback, and benefit from personalized learning. Tools like ChatGPT \cite{Zhang2023} and Claude \cite{AlAli2024} exemplify this shift by providing real-time assistance that enhances the learning experience. Retrieval-Augmented Generation (RAG) \cite{Chen2020} is often used to add context to the large language model (LLM) based on course materials, instructor instructions, and conversation history. It is a proven approach that allows AI systems to offer better explanations and context-aware tutoring. 

However, the accuracy of AI-generated content is not guaranteed, particularly when dealing with novel problems or edge cases, complicating the learning process \cite{chan2019,ref24}. Moreover, GenAI introduces concerns about the authenticity of student work. How can educators ensure that the solutions reflect the student’s understanding rather than the AI’s output? Educators must therefore redefine assessment frameworks to evaluate both AI-supported and independent work fairly, which requires a nuanced understanding of the interaction between technology and pedagogy \cite{ref34,ref26}.

In the context of computer science education, GenAI tools provide exciting opportunities for enhancing efficiency and creativity in teaching foundational skills such as coding, computational thinking, and problem-solving. However, the accuracy of AI-generated solutions remains a significant concern, as AI hallucinations can lead to incorrect or irrelevant information, but students may lack the knowledge to fact-check AI responses. GenAI also raises concerns about authenticity, where it is unclear whether students are truly learning to solve problems independently or relying on AI-generated solutions. Additionally, the auto-generation of assignments complicates the learning environment and raises questions about the depth of student engagement \cite{ref23,ref1}. These emerging shifts in AI use in computer science teaching and learning pose challenges for traditional assessment methods like exams or project-based evaluations, which may no longer suffice in measuring student comprehension and skill development \cite{ref30,ref10}. From a teacher’s perspective, AI-generated content can seamlessly blend with human effort, raising concerns about the authenticity of student work. This blending also complicates the verification of accuracy, particularly in complex programming scenarios, requiring new levels of scrutiny from educators \cite{ref6,ref28}. As a result, designing assessment frameworks that reflect individual student skills while accounting for AI support becomes an increasingly complex task \cite{ref29,ref14}.

This survey synthesizes recent research on GenAI in computer science education, focusing on the themes of accuracy, authenticity, and assessment. While many studies have explored these themes individually, there is a notable gap in the literature regarding their interconnection and integration. By examining how these three issues interact, this paper seeks to provide a better understanding of how GenAI can transform teaching and learning in computer science. Additionally, this study summarizes areas for further research \cite{ref5,ref24}.

While this study focuses on accuracy, authenticity, and assessment, these themes are part of a broader and ongoing discussion about GenAI’s impact on education. We aim to explore the relationship between AI technologies and the evolving educational landscape, particularly within computer science. The goal is to highlight both the transformative potential and the significant challenges these technologies introduce into teaching and learning processes \cite{ref2,ref7}, fostering a more informed approach to their integration in computer science curricula \cite{ref13,ref15}. Other critical considerations, such as ethical implications, student engagement, and the development of new technical literacies, also demand attention. We propose these as areas for future work.

\section{Literature Review}

We reviewed 52 papers published between 2019 and 2024, sourced from academic databases such as IEEE Xplore, ACM Digital Library, and SpringerLink. These studies were categorized into three primary themes: accuracy, authenticity, and assessment. Most of the papers focused on accuracy, particularly addressing AI hallucinations and bias, while fewer discussed the challenges of assessment in AI-driven educational settings.

\subsection{Accuracy}
The accuracy of generative AI (GenAI) in computer science education has garnered significant attention as educators and researchers strove to incorporate AI technologies into learning environments. Accuracy is crucial for generating reliable content, offering effective educational feedback, and supporting student learning. However, there were significant concerns about AI hallucinations (i.e., the generation of factually incorrect or nonsensical information), error propagation (i.e., when initial mistakes persist and compound in subsequent outputs), bias, and the delicate balance between creativity and precision. 

\subsubsection{Implications of AI Hallucination and Error Propagation for Accuracy}
AI hallucinations and error propagation were central issues in the adoption of GenAI models in CS education, particularly when AI-generated outputs accumulated inaccuracies over time. AI hallucinations referred to instances where AI systems produced incorrect or nonsensical information with high confidence. Wu et al.~\cite{wu2024} highlighted the prevalence of hallucinations, reporting that 53\% of the errors in their study were classified as hallucinations, which could mislead both students and educators.

Hallucinations compromised the reliability of AI-generated content, leading to concerns about student learning outcomes. For instance, Butler~\cite{butler2024} noted that hallucinations detracted from the overall quality of AI-generated reports, potentially eroding trust in AI tools as educational aids. These errors could accumulate over time, propagating through multiple layers of content generation and significantly affecting the educational process. 

Prather et al.~\cite{ref27} reinforced these concerns, particularly for novice learners, who might not yet have had the skills to critically assess AI-generated outputs. Lin et al.~\cite{ref19} further noted that models trained on biased or incomplete datasets often generated erroneous content, complicating the learning process. Becker et al.~\cite{ref3} offered a counterpoint, suggesting that robust feedback systems could mitigate the effects of error propagation by allowing students to cross-reference AI outputs with reliable sources.

Sharun~\cite{sharun2023} proposed that improving the diversity and quality of training datasets was a crucial step toward reducing hallucinations and enhancing the accuracy of AI systems. Additionally, Aditya~\cite{aditya2024} emphasized the role of human-in-the-loop (HITL) systems, where human oversight was integrated into the AI output generation process to provide real-time corrections and ensure accuracy. This approach aligned with the need to reduce error propagation and mitigate the negative impact of hallucinations in educational settings.

\subsubsection{Challenges and Mitigation Strategies of Bias}
Bias in AI-generated content was another critical issue affecting accuracy. Mahaini~\cite{ref22} discussed how GenAI tools, while adept at solving complex programming tasks, often introduced biases that could distort accuracy, particularly in assessments. This bias could disproportionately impact students from diverse backgrounds, as biased models might overlook cultural or contextual differences in problem-solving approaches. Emenike and Emenike~\cite{ref8} argued that reliance on biased AI-generated assessments could exacerbate educational inequalities.

To address these challenges, Gupta et al.~\cite{ref12} proposed hybrid models that combined human oversight with AI-generated content to enhance accuracy and mitigate bias. Walter~\cite{ref31} emphasized the importance of AI literacy and critical thinking, enabling students and educators to recognize and correct biased outputs. These strategies underscored the essential role of human involvement in addressing the challenges posed by bias in AI-generated content.

\subsubsection{Challenges in Measuring and Evaluating Accuracy}
Measuring and evaluating the accuracy of AI-generated content posed ongoing challenges. Francisco and Silva~\cite{ref9} pointed out that current accuracy metrics often failed to account for the contextual nature of educational tasks, where solutions might not be strictly right or wrong. Similarly, Chan and Hu~\cite{ref4} suggested that traditional metrics focusing solely on error rates overlooked the broader educational value of AI-generated content. Becker et al.~\cite{ref3} proposed a holistic approach to accuracy, where metrics encompassed not only the correctness of the output but also how well AI-generated content supported learning goals. Liu~\cite{ref21} proposed hybrid metrics that combined traditional accuracy measures with qualitative assessments of student engagement and problem-solving skills, offering a more comprehensive picture of AI's educational effectiveness.

\subsubsection{Creativity vs. Accuracy: A Careful Balance}
The trade-off between creativity and accuracy was another common concern. Pack and Maloney~\cite{ref25} noted that while GenAI was celebrated for its creative solutions, this creativity often came at the expense of accuracy. In computer science education, where precise solutions were paramount, this trade-off could hinder learning. Francisco and Silva~\cite{ref9} advocated for a balance between creativity and accuracy, suggesting the implementation of context-aware AI systems that adjusted their outputs based on the task. For instance, while creative solutions might be encouraged in brainstorming exercises, accuracy should take precedence in tasks requiring code generation.

Zastudil et al.~\cite{ref36} cautioned that an overemphasis on accuracy might stifle creativity, particularly in fields like computing where innovation was crucial. They argued that students should be encouraged to experiment with AI-generated outputs, even if these were not always accurate, as such experimentation fostered deeper engagement and problem-solving skills. This highlighted the ongoing discussion of promoting creativity while maintaining accuracy in AI-generated content.

\subsubsection{Proposed Solutions for Improving Accuracy}
Several strategies had been proposed to enhance the accuracy of GenAI models in computer science education. Kung et al.~\cite{ref16} suggested that improving the diversity and quality of training datasets could significantly enhance model accuracy, particularly in coding tasks where precision was essential. Incorporating diverse datasets helped AI models generate outputs that were less prone to errors and biases. Liu~\cite{ref21} advocated for employing reinforcement learning techniques to fine-tune AI models for educational purposes, allowing models to learn from their mistakes and improve accuracy over time. This aligned with the findings of Lee and Song~\cite{ref18}, who emphasized the role of iterative feedback in refining AI-generated outputs, particularly in programming tasks.

Human-AI interaction was crucial for improving accuracy. Mahaini~\cite{ref22} argued that while AI tools could generate highly accurate content, their effectiveness was enhanced when used in conjunction with human oversight. Educators could provide contextual understanding and correct errors, thereby improving the overall quality of AI-generated outputs. Walter~\cite{ref31} emphasized the importance of AI literacy among educators and students, suggesting that informed users could guide AI inputs and interpret outputs more effectively, thus enhancing accuracy. However, Lee and Song~\cite{ref18} cautioned against over-reliance on human intervention, noting that AI models should be designed to operate as accurately as possible independently, to avoid overburdening educators.

\subsubsection{Future Research Directions}
Despite advancements in improving accuracy, several research gaps persisted. One significant gap was the lack of longitudinal studies examining the long-term impact of AI-generated inaccuracies on student learning outcomes. Chan and Hu~\cite{ref4} noted that while short-term studies demonstrated the benefits of GenAI in education, little was known about how inaccuracies in AI-generated content affected students' academic performance over time. Additionally, the development of bias detection and mitigation techniques for AI-generated assessments remained underexplored. While some strategies, such as those proposed by Gupta et al.~\cite{ref12}, aimed to reduce bias, comprehensive frameworks were needed to integrate these techniques into educational systems effectively.

The balance between creativity and accuracy also required further investigation. Adaptive AI models that dynamically adjusted between fostering innovation and maintaining precision based on task requirements had been proposed, but empirical work was needed to refine these approaches. Furthermore, developing nuanced metrics to evaluate accuracy beyond simple error rates would provide educators with better insights into how AI-generated content impacted student learning.

\subsubsection{Summary of GenAI Accuracy}
Table~\ref{tab:keyfindings_accuracy} summarizes key findings on the theme of accuracy in generative AI for computer science education.

\captionsetup[table]{justification=centering, singlelinecheck=false, skip=3pt} % Center and move the caption slightly higher

\renewcommand{\arraystretch}{1.4} % Adjusts vertical padding for the table
\setlength{\tabcolsep}{10pt} % Adjusts horizontal padding for each cell

\begin{footnotesize} % Change to scriptsize or tiny for smaller font
\begin{longtable}{|p{0.3\textwidth}|p{0.1\textwidth}|p{0.3\textwidth}|}
\caption{Key Findings on Accuracy in Generative AI for Computer Science Education}
\label{tab:keyfindings_accuracy} \\
\hline
\multicolumn{1}{|c|}{\textbf{Challenges}} & 
\multicolumn{1}{c|}{\textbf{Ref. Cited}} & 
\multicolumn{1}{c|}{\textbf{Opportunities}} \\ \hline  
\endfirsthead

\multicolumn{3}{c}{{\tablename\ \thetable{} -- continued from previous page}} \\ \hline
\multicolumn{1}{|c|}{\textbf{Challenges}} & 
\multicolumn{1}{c|}{\textbf{Ref. Cited}} & 
\multicolumn{1}{c|}{\textbf{Opportunities}} \\ \hline
\endhead

\hline \multicolumn{3}{|r|}{{Continued on next page}} \\ \hline
\endfoot

\hline
\endlastfoot

AI hallucinations can mislead students and educators, undermining content reliability. 
& [21,22,23,26] &
Conduct long-term studies on the cumulative effects of AI hallucinations on learning outcomes. \\ \hline  

Error propagation amplifies inaccuracies, especially in AI-assisted programming. 
& [24,23,25,27] &
Developing real-time mitigation strategies for error propagation, particularly benefiting novice learners. \\ \hline  

Bias in AI affects the accuracy of assessments, particularly for diverse student groups. 
& [28,29,30] &
Development frameworks for bias detection and mitigation in AI-driven assessments. \\ \hline  

Creativity vs. accuracy trade-off in AI-generated content hinders precision in coding tasks. 
& [32,33,34] &
Explore adaptive models that dynamically balance creativity and accuracy in AI systems. \\ \hline  

Human-in-the-loop (HITL) systems improve AI accuracy and support real-time corrections. 
& [27,30,31,25] &
Scaling HITL models for large educational contexts. \\ \hline  

AI struggles with adapting to educational contexts where correctness is subjective, highlighting the need for holistic accuracy measures. 
& [33,34,32] &
Developing AI metrics that better capture the contextual nature of educational tasks. \\ \hline  

\end{longtable}
\end{footnotesize}

\subsection{Authenticity}
\label{sec:authenticity}

The theme of authenticity in the context of GenAI within computer science education has gained prominence as AI-generated outputs become increasingly prevalent. Key concerns revolve around the originality of student work, the ethical implications of AI-generated content, and the challenges educators face in distinguishing between human-authored and AI-generated submissions. This review synthesizes key findings from the literature, exploring how the integration of GenAI affects the authenticity of student work in educational settings.

\subsubsection{AI-Generated Content and Challenges to Student Authorship}
A primary concern is the risk of students presenting AI-generated content as their own, raising issues of authorship and academic integrity. Prather et al.~\cite{ref3} argued that as tools like ChatGPT evolved, differentiating between human-authored and AI-generated work became increasingly difficult, particularly in coding assignments where correctness is paramount. Becker et al.~\cite{ref3} suggested that the accessibility of AI-generated code diminished the authenticity of student submissions. Lin et al.~\cite{ref19} noted that students may unintentionally incorporate AI-generated outputs into their work, blurring the line between original thought and AI assistance, which could hinder the development of critical thinking and problem-solving skills. Conversely, Zastudil et al.~\cite{ref36} acknowledged that AI tools could enhance learning by providing immediate feedback but emphasized the need for clear guidelines to prevent misuse.

\subsubsection{Ethical Concerns Surrounding AI-Generated Content}
Ethical considerations regarding AI-generated content are central to discussions about authenticity. Emenike and Emenike~\cite{ref8} highlighted that AI-generated content, particularly in written assignments, posed significant challenges for educators in assessing genuine student engagement. Over-reliance on AI tools may have diminished students' ability to produce original work. Gupta et al.~\cite{ref12} emphasized the difficulty of detecting AI-generated text, noting that traditional plagiarism detection tools often failed to recognize it. They advocated for institutions to invest in AI-detection technologies to uphold academic integrity. Francisco and Silva~\cite{ref9} adopted a more optimistic perspective, arguing that AI could be used ethically if students were transparent about its usage, framing AI as a learning aid rather than a substitute for original work.

\subsubsection{The Sociotechnical Perspective on AI Authenticity}

Addressing the ethical implications of AI-generated content and authorship in education requires a sociotechnical perspective, where the interaction between social dynamics and technological tools is key to understanding authenticity issues. In computer science education, AI tools like ChatGPT and GitHub Copilot present unique challenges to the authenticity of student work, raising concerns about academic integrity and necessitating a holistic approach.

The authenticity of student work is increasingly questioned as AI tools generate text and code. Reliance on AI tools like ChatGPT can lead to reduced emphasis on original thought, with some researchers warning that such tools might affect creativity and critical thinking skills~\cite{Er2023,Students2023}. Educators must design assignments that promote deeper engagement and critical thinking, moving beyond tasks easily completed by AI~\cite{Er2023}.

The ethical implications of authorship are complex, as AI blurs the lines between student-authored and AI-generated content. This raises concerns about plagiarism and academic integrity~\cite{Crawford2023,Foltunek2023}. Clear guidelines are essential to ensure students understand how to responsibly use AI, including educating them about AI's limitations and the consequences of over-reliance~\cite{Perera2023,Er2023,Lawrie2023}.

A sociotechnical approach can help educators distinguish between human and AI-generated content by leveraging AI's capabilities while preserving student work integrity. Assessment methods like oral presentations and reflective essays can ensure students demonstrate personal understanding~\cite{Thanh2023,Yu2023}. Integrating AI literacy into curricula can also empower students to use these tools responsibly~\cite{Delcker2024}.

Transparency and fairness are critical as AI tools become more prevalent in education. Students with limited access to technology may face disadvantages~\cite{Williams2022,Kuleto2021}, so educators must provide resources and training to ensure equitable access to AI technologies~\cite{Williams2022}. Ethical considerations must guide AI’s implementation, preparing students to navigate technology responsibly while maintaining academic integrity~\cite{Halaweh2023}.

\subsubsection{Summary of GenAI Authenticity}
Table~\ref{tab:keyfindings_authenticity} summarizes key findings on the theme of Authenticity in generative AI for computer science education.

\captionsetup[table]{justification=centering, singlelinecheck=false, skip=3pt} % Center and move the caption slightly higher

\renewcommand{\arraystretch}{1.4} % Adjusts vertical padding for the table
\setlength{\tabcolsep}{6pt} % Reduce horizontal padding for each cell

\begin{footnotesize} % Change to scriptsize or tiny for smaller font
\begin{longtable}{|p{0.35\textwidth}|p{0.12\textwidth}|p{0.35\textwidth}|} % Adjusted reference column width
\caption{Key Findings on Authenticity in Generative AI for Computer Science Education}
\label{tab:keyfindings_authenticity} \\
\hline
\multicolumn{1}{|c|}{\textbf{Challenges}} & 
\multicolumn{1}{c|}{\textbf{Ref. Cited}} & 
\multicolumn{1}{c|}{\textbf{Opportunities}} \\ \hline  
\endfirsthead

\multicolumn{3}{c}{{\tablename\ \thetable{} -- continued from previous page}} \\ \hline
\multicolumn{1}{|c|}{\textbf{Challenges}} & 
\multicolumn{1}{c|}{\textbf{Ref. Cited}} & 
\multicolumn{1}{c|}{\textbf{Opportunities}} \\ \hline
\endhead

\hline \multicolumn{3}{|r|}{{Continued on next page}} \\ \hline
\endfoot

\hline
\endlastfoot

AI blurs the line between student-authored and AI-generated work, raising academic integrity concerns. 
& [23, 25, 24, 43, 39, 44] &
Develop clear policies to distinguish AI-assisted work from student-authored work. \\ \hline  

Ethical concerns arise due to the difficulty in detecting AI-generated content in student submissions. 
& [29, 30, 32, 41, 42] &
Create better plagiarism detection tools and institutional guidelines to ensure academic integrity. \\ \hline  

Sociotechnical approaches can help preserve authenticity in student work. 
& [47, 46, 45, 48, 49] &
Explore a combination of sociotechnical approaches to promote authenticity. \\ \hline  

Transparent use of AI tools, combined with AI literacy education, can help mitigate ethical concerns while preserving student engagement and creativity. 
& [32, 30, 36] &
Opportunity for more empirical studies on the effectiveness of AI literacy programs. \\ \hline  

Responsible use of AI supports learning without compromising authenticity. 
& [36, 23, 24] &
Explore the balance between ethical AI usage and maintaining originality in student work. \\ \hline  

\end{longtable}
\end{footnotesize}

\subsection{Assessment}
The integration of GenAI in education has significantly transformed assessment practices, particularly within computer science. AI tools are increasingly employed for grading, providing feedback, and evaluating student performance, streamlining the assessment process while raising concerns about fairness, reliability, and objectivity. This section reviews key discussions on the role of AI in assessment, offering a thematic analysis of its impact on computer science education.

\subsubsection{AI-Driven Assessment Tools: Grading and Feedback}
AI-driven grading tools have gained attention for their ability to automate grading and deliver timely feedback. Lin et al.~\cite{ref19} noted that these tools could provide real-time feedback on programming tasks, enabling students to correct mistakes as they worked, which was particularly beneficial in large classrooms where human grading was challenging. Prather et al.~\cite{ref27} emphasized that AI grading systems evaluated code for correctness and efficiency, alleviating the burden on educators during routine assessments.

However, Becker et al.~\cite{ref3} cautioned that AI-driven grading tools might struggle with complex or creative problem-solving tasks that required deeper contextual understanding. Francisco and Silva~\cite{ref9} argued that while AI effectively assessed straightforward coding tasks, it fell short in evaluating innovative algorithms or creative solutions. In contrast, Walter~\cite{ref31} suggested that AI assessments enhanced objectivity and consistency by minimizing human biases, such as favoritism and fatigue.

\subsubsection{Fairness, Objectivity, and Ethical Challenges in AI-Driven Assessments}
A significant debate in the literature centered on the fairness and objectivity of AI-driven assessments. Zastudil et al.~\cite{ref36} expressed concerns that AI tools could perpetuate biases if trained on skewed data, leading to unfair evaluations. For instance, AI systems might favor specific coding styles based on their training data. Kung et al.~\cite{ref16} further highlighted that students from diverse backgrounds might face unfair assessments if AI tools were not trained on inclusive datasets.

Prather et al.~\cite{ref27} advocated for a hybrid approach that combined AI systems with human assessments to ensure both objectivity and nuanced judgment. This model balanced the strengths of AI evaluations with the contextual understanding human graders provided. However, Mahaini~\cite{ref22} emphasized that human oversight was crucial, particularly for summative assessments where creativity and critical thinking had to be evaluated.

\subsubsection{Evaluation of AI-Generated Content: Metrics and Limitations}
The limitations of traditional grading metrics in assessing AI-generated content were a major concern. Mahaini~\cite{ref22} noted that while AI systems excelled at evaluating code correctness, they struggled with assessing creativity and originality—key learning outcomes in computer science. González-Calatayud et al.~\cite{ref11} argued that metrics focused solely on correctness and efficiency failed to capture essential critical thinking and problem-solving skills.

Several authors, including Liu et al.~\cite{ref21}, advocated for new assessment frameworks that evaluated both technical and creative aspects of student work. Becker et al.~\cite{ref3} suggested training AI systems to recognize diverse coding styles and problem-solving approaches to avoid penalizing students for innovative thinking.

\subsubsection{Formative and Summative Assessment: The Role of AI in Supporting Learning}
AI tools could enhance both formative and summative assessments. Lee~\cite{ref17} highlighted that AI could provide formative feedback throughout the learning process, helping students identify areas for improvement before final evaluations, thereby reducing pressure on summative assessments. However, Mahaini~\cite{ref22} cautioned against relying solely on AI for summative assessments, stressing the necessity of human judgment in evaluating creativity and critical thinking. Pack and Maloney~\cite{ref25} also emphasized the importance of human involvement in assessing complex tasks, such as innovative coding projects, which might exceed the capabilities of AI-driven assessments.

\subsubsection{Human-AI Collaboration in Assessments}
There was a consensus that AI should complement, not replace, human involvement in assessments. Mahaini~\cite{ref22} and Becker et al.~\cite{ref3} asserted that while AI could efficiently handle routine grading tasks, human oversight was essential for more complex evaluations, such as coding projects and creative problem-solving exercises. Francisco and Silva~\cite{ref9} proposed a hybrid assessment model where AI conducted an initial evaluation, followed by human review to ensure that the assessment accurately reflected the student's understanding and creativity. This model combined AI's efficiency with human insight, facilitating a more holistic evaluation process.

\subsubsection{Research Gaps and Future Directions}
Despite the growing body of literature on AI-driven assessments, several gaps remained. Notably, there was a lack of long-term studies examining the impact of AI-driven assessments on student learning outcomes. While short-term studies demonstrated the benefits of AI in providing timely feedback, limited research existed on how these tools affected students' development of critical thinking and problem-solving skills over time~\cite{ref27,ref19}.

Another significant gap was the impact of AI-generated feedback on student motivation. Gupta et al.~\cite{ref12} noted that while AI could provide instant feedback, its effectiveness in motivating students compared to human feedback remained unclear. Future research should explore how students perceive and respond to AI-generated feedback.

Additionally, more research was needed to ensure fairness, transparency, and objectivity in AI-driven assessments. Strategies for mitigating bias in AI systems, as discussed by Zastudil et al.~\cite{ref36} and Kung et al.~\cite{ref16}, required further exploration. Addressing these research gaps was essential to ensure that AI-driven assessments enhanced the learning process while upholding educational integrity.

\subsubsection{Summary of GenAI Assessment}
Table~\ref{tab:keyfindings_assessment} summarizes key findings on the theme of Assessment in generative AI for computer science education.
\captionsetup[table]{justification=centering, singlelinecheck=false, skip=3pt} % Center and move the caption slightly higher

\renewcommand{\arraystretch}{1.4} % Adjusts vertical padding for the table
\setlength{\tabcolsep}{4pt} % Further reduce horizontal padding for each cell

\begin{footnotesize} % Change to scriptsize or tiny for smaller font
\begin{longtable}{|p{0.35\textwidth}|p{0.15\textwidth}|p{0.35\textwidth}|} % Increased reference column width
\caption{Key Findings on Assessment in Generative AI for Computer Science Education}
\label{tab:keyfindings_assessment} \\
\hline
\multicolumn{1}{|c|}{\textbf{Challenges}} & 
\multicolumn{1}{c|}{\textbf{Ref. Cited}} & 
\multicolumn{1}{c|}{\textbf{Opportunities}} \\ \hline  
\endfirsthead

\hline
\endlastfoot

AI-driven grading tools improve efficiency and provide timely feedback in large classes. 
& [19, 27, 31, 22] &
Incorporate AI for evaluating creativity and problem-solving with human oversight. \\ \hline  

AI grading enhances objectivity and consistency, reducing human biases such as favoritism. 
& [31, 9] &
Improve AI training with diverse datasets to minimize bias in assessments. \\ \hline  

Hybrid AI-human evaluation balances efficiency with nuanced judgment for complex tasks. 
& [22, 27, 3, 22] &
Develop scalable hybrid models for broader use in diverse educational contexts. \\ \hline  

New frameworks are needed to evaluate both technical and creative aspects of student work. 
& [22, 21, 11] &
Create comprehensive frameworks for assessing creativity and accuracy. \\ \hline  

AI enhances formative assessments by providing real-time feedback to guide learning. 
& [17, 25, 9, 12] &
Study long-term effects of AI-driven formative feedback on student learning outcomes. \\ \hline  

Human oversight is essential for summative assessments requiring creativity and critical thinking. 
& [22, 25, 3, 22] &
Explore optimal combinations of AI and human input in summative assessments. \\ \hline  

\end{longtable}
\end{footnotesize}

\section{Discussion}
This review highlights the significant opportunities and challenges posed by the integration of generative AI in computer science education, particularly in terms of accuracy, authenticity, and assessment. While AI tools such as ChatGPT and GitHub Copilot offer benefits like increased efficiency, real-time feedback, and creative assistance, they also introduce critical issues like AI hallucinations, error propagation, and biases. Ensuring the accuracy of AI-generated content remains a significant challenge, particularly for novice learners, where errors and biases could adversely affect learning outcomes. Moreover, the authenticity of student work is jeopardized by the blurring of lines between AI-assisted and student-generated content, raising concerns about academic integrity. To address these challenges, educators must implement hybrid assessment models that combine human judgment with AI tools, promote AI literacy, and develop clear policies on the ethical use of AI. Future research should focus on refining AI systems to balance creativity with precision, enhance bias detection frameworks, and study the long-term effects of AI-generated inaccuracies on student learning.

\subsubsection{Research Questions for Future Studies}
To address these gaps, future research could focus on the following questions:
\begin{itemize}
    \item How does long-term exposure to AI-generated inaccuracies affect students' problem-solving abilities and coding skills?
    \item What strategies can be developed to more effectively detect and mitigate bias in AI-generated educational content?
    \item How can adaptive AI models be designed to balance the need for creativity and accuracy in different educational contexts?
    \item What are the ethical implications of using AI detection tools in academic assessments, and how can institutions ensure that these tools are used transparently and fairly?
\end{itemize}

\bibliographystyle{unsrt}
\bibliography{references}
\end{document}